\documentclass[reprint,a4paper,amsmath,amssymb,floatfix,showkeys,showpacs]{revtex4-1}

\usepackage{graphicx}
\usepackage{upgreek}
\usepackage{textcomp}

\begin{document}

\title{Measurement of the gas-flow reduction factor of the KATRIN \mbox{DPS2-F} differential pumping section}

\author{S. Luki\'c}
\email{slukic@vinca.rs}
\altaffiliation{Present address: VIN\v{C}A Institute of Nuclear Sciences, Belgrade, Serbia}
\author{B. Bornschein}
\author{L. Bornschein}
\author{G. Drexlin}
\author{A. Kosmider}
\author{K. Schl\"osser}
\author{A. Windberger}
\altaffiliation{Present address: Max-Planck-Institut f\"ur Kernphysik, Heidelberg, Germany}
\affiliation{Karlsruhe Institute of Technology, Karlsruhe, Germany}

\date{\today}

\begin{abstract}
The gas-flow reduction factor of the second forward Differential Pumping Section (\mbox{DPS2-F}) for the KATRIN experiment was determined using a dedicated vacuum-measurement setup and by detailed molecular-flow simulation of the \mbox{DPS2-F} beam tube and of the measurement apparatus. In the measurement, non-radioactive test gases deuterium, helium, neon, argon and krypton were used, the input gas flow was provided by a commercial mass-flow controller, and the output flow was measured using a residual gas analyzer, in order to distinguish it from the outgassing background. The measured reduction factor with the empty beam tube at room temperature for gases with mass 4 is $1.8(4) \times 10^{4}$, which is in excellent agreement with the simulated value of $1.6 \times 10^{4}$. The simulated reduction factor for tritium, based on the interpolated value for the capture factor at the turbo-molecular pump inlet flange is $2.5 \times 10^{4}$. The difference with respect to the design value of $1 \times 10^5$ is due to the modifications in the beam tube geometry since the initial design, and can be partly recovered by reduction of the effective beam tube diameter.
\end{abstract}

\maketitle

\section{Introduction}
\label{}

The KArlsruhe TRItium Neutrino experiment (KATRIN), currently under construction at Campus North of the Karlsruhe Institute of Technology (KIT), will measure the electron neutrino mass in a model-independent way by measuring the shape of the energy spectrum of electrons from tritium $\beta$-decay \cite{KATRIN04}.

The experimental set-up of KATRIN has already been described in several publications \cite{KATRIN04, Dre05, Bonn08}. The $\beta$-electrons in the KATRIN experiment will originate from the 10 m long Windowless Gaseous Tritium Source (WGTS), where a stable density profile will be maintained by the continuous injection of tritium gas at the center of the vacuum tube with a throughput of $1.8 \text{\ mbar l/s}$, and by continuous pumping of the gas at both ends in the first stages of the differential pumping section (DPS1). In the forward direction, $\beta$-electrons from the WGTS will be transported by magnetic fields of up to 5.6~T towards the high-resolution electrostatic spectrometer, where their kinetic energy will be analyzed. The electron-transporting magnetic fields are created by superconducting solenoids installed at each section of the beamline. Along the beamline, the tritium flow is further reduced in the second differential pumping section (\mbox{DPS2-F}), followed by the Cryogenic Pumping Section (CPS). Before the beamline reaches the spectrometer section, the tritium flow has to be reduced by 14 orders of magnitude, in order to keep the tritium-induced background below 1~mHz \cite{KATRIN04}.

\begin{figure*}
\centering
\includegraphics[width=\textwidth]{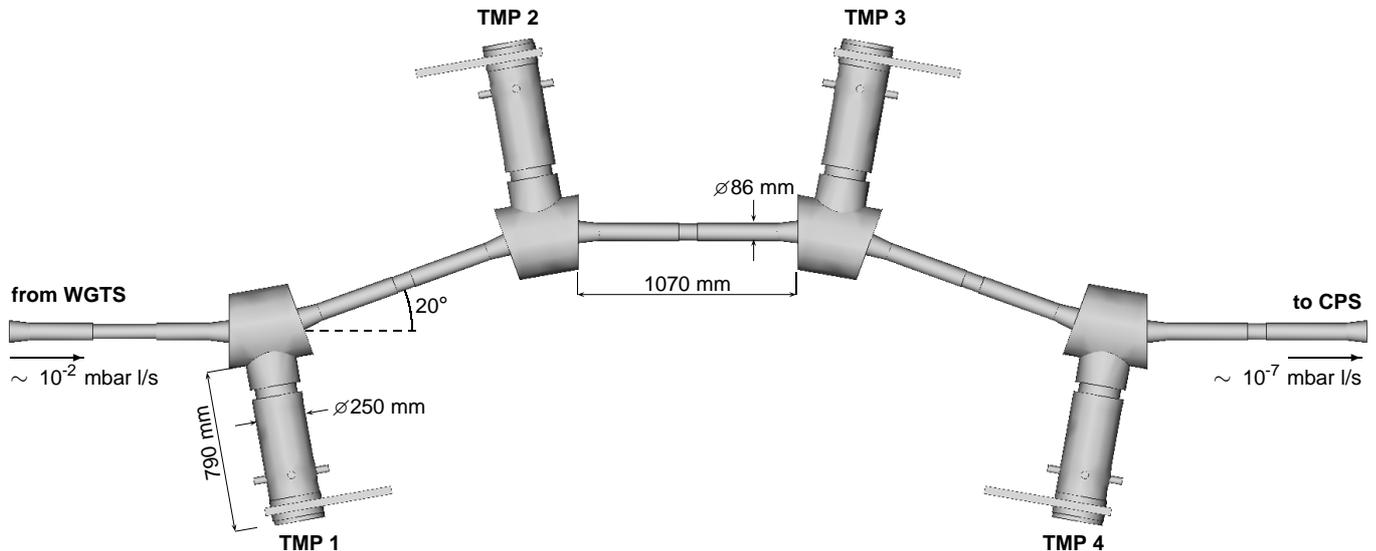}
\caption{\label{DPS2}Inner geometry of the \mbox{DPS2-F} beam tube.}
\end{figure*}

The \mbox{DPS2-F} represents the last stage of turbo-molecular pumping in the transport section. As there is a strong effect of molecular beaming at the outlet of WGTS \cite{Mal09}, the successive segments of the \mbox{DPS2-F} beam tube are inclined by 20\textdegree{} with respect to each other as shown in Fig. \ref{DPS2}. Each of the five beam-tube segments has a total length of 1070~mm, and consists of two equal straight sections with the inner diameter 86~mm and one bellow between the straight sections with the length of 90~mm and the minimum diameter of 81~mm. Between the successive segments, four pumping ports are installed. Each pumping port consists of a cyllindrical chamber cut to trapezoidal shape in order to accomodate the 20\textdegree{} inclination between the tube segments, a 670~mm long duct with 250~mm diameter, a gate valve and a Leybold MAG~W~2800 turbo-molecular pump (TMP; not shown in the figure). The minimum length of the pumping ducts is dictated by the necessity to keep the TMP outside of the strong magnetic field of the \mbox{DPS2-F} magnets as well as to reduce the thermal radiation load from the TMP which is at room temperature. For assembly reasons, the pumping duct features an inner flange, seen as a short reduction near the chamber. On each side of the inner flange, a 82~mm long bellow provides insulation against thermal conductance between the TMP and the beam tube which is nitrogen cooled. 

The gas-flow retention capability of the \mbox{DPS2-F} is of high importance for keeping the spectrometers essentially tritium free, as well as for the required frequency of the CPS regeneration \cite{KATRIN04}. The retention capability is customarily expressed in terms of the gas-flow reduction factor, defined as the ratio of the gas-flow at the \mbox{DPS2-F} inlet to that at its outlet. The reduction factor has been extensively studied in the design phase by calculation and gas-flow simulation \cite{Mal09, Bonn03, Luo06}. Results of these estimates depend strongly on the input parameters regarding the \mbox{DPS2-F} geometry and on the boundary conditions that were not all precisely known at that time. 

A dedicated vacuum-measurement setup was built and calibrated to measure the reduction factor. At the time of this experiment, tritium could not be used because the \mbox{DPS2-F} was not yet integrated into the Tritium Laboratory (TLK). Therefore, deuterium, helium, neon, argon and krypton were used instead, as test gases. As shown in \cite{Luo06}, the reduction factor of the \mbox{DPS2-F} depends on the capture factor at each of the flanges connecting the TMP, which in turn depends on the molecular mass of the gas in question \cite{Mal07}. The capture factor at a given boundary surface is defined as the ratio of the number of molecules pumped at the surface to the number of molecules impinging on the surface in the same time interval.\footnote{Specifically in the case of TMP, the common term in vacuum technology is the ``Ho-factor''. For the sake of uniformity, in this work the term ``capture factor'' is used in all cases, regardless whether the boundary surface is defined at the inlet of a TMP or at some other boundary between subsystems, because all surfaces where capture factors are defined are treated in an equivalent way in the simulations presented here.} 
According to the molecular dynamics, the capture factor $\alpha$ relates to the effective pumping speed via the relation $S(M) = \frac{1}{4} \alpha(M) \bar{v}(M) A$, where $\bar{v}(M)$ is the mean molecular speed, and $A$ is the area of the pumping surface. As shown in \cite{Mal07}, characterization of TMP by the capture factor leads to a simple expression as a function of the molecular mass of the pumped gas, allowing interpolation to gases for which the pumping speed was not measured. This expression is used here to estimate the \mbox{DPS2-F} reduction factor for tritium by interpolation of the TMP capture factors, using measurement data for the test gases and gas-flow simulations. 

Another issue of importance for this experiment is the dependence of the reduction factor on the capture factor seen by molecules at the outlet of the \mbox{DPS2-F}. This dependence was determined using gas-flow simulations as well, and the appropriate correction was applied to the measurement results.

In the following section the gas-flow measurement setup is described. In Sec. \ref{sec-analysis} the analysis of the measurement using gas-flow simulations is discussed. The calibration procedure is described in Sec. \ref{sec-calibration}. Results are presented and compared to the value estimated for the original design geometry in Sec. \ref{sec-results}, and the summary is given in Sec. \ref{sec-summary}.

\section{Measurement setup}
\label{sec-setup}

\begin{figure}
\centering
\includegraphics{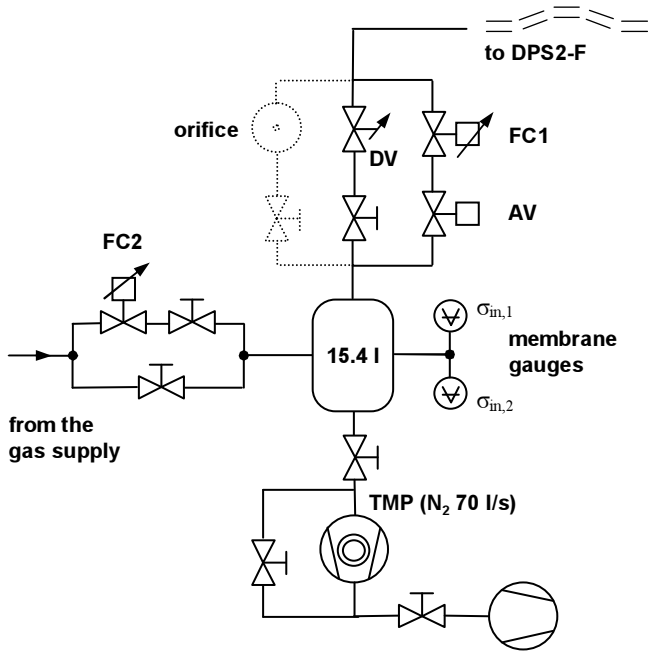}
\caption{Injection system with the 15~l pressure buffer, the mass-flow controllers (FC) and the manual dosing valve (DV). The calibration orifice (shaded) was installed only in the calibration phase.}
\label{injection}
\end{figure}

Stable throughput of the test gas at the \mbox{DPS2-F} inlet, corresponding to the expected throughput from the WGTS, was provided by the injection system schematically shown in Fig. \ref{injection}. The injection system comprises a pressure buffer with a volume of 15.4~l and the MKS M200 mass-flow controller (FC1) with full scale of 1.1~sccm (about 0.02~mbar~l/s at room temperature) for gas injection into the \mbox{DPS2-F}. The pressure buffer is fed with selected gas of purity of at least 99.99\% from a gas-supply cabinet. For pressure monitoring, two MKS Baratron\textsuperscript\textregistered capaticance-diaphragm pressure transducers, one with a full scale of 1.33~mbar ($\upsigma_{\text{in},1}$), and another with a full scale of 1333~mbar ($\upsigma_\text{in,2}$) are installed in the system. Another MKS M200 mass-flow controller (FC2) with full scale of 2~sccm is used when necessary to maintain the pressure in the buffer and is automatically controlled according to the pressure signal from the transducer $\upsigma_\text{in,2}$. A manual dosing valve was installed in a parallel output line. In the calibration phase of the measurement system (see Sec. \ref{sec-calibration}), an orifice was installed in a third output line. 

\begin{figure}
\centering
\includegraphics{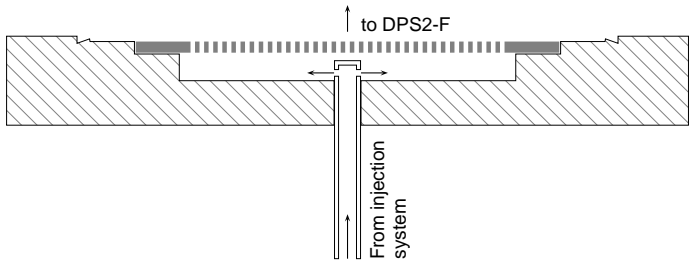}
\caption{Gas injection chamber}
\label{inlet}
\end{figure}

In order to provide similar gas-flow distribution both in the calibration and in the measurement phase, as well as to roughly simulate the KATRIN gas-flow conditions, the outlet of the injection system was equipped with the injection chamber shown in Fig. \ref{inlet}. The chamber was built by milling a cyllindrical cavity in a standard CF-100 blind flange and covering it with a stainless-steel plate with ca. 1400 drilled openings, each with a diameter of 1~mm. The gas was supplied via a 6~mm tube welded into the flange from the opposite side. The tube end had lateral openings in order to distribute the gas in the chamber as evenly as possible. 

\begin{figure}
\centering
\includegraphics{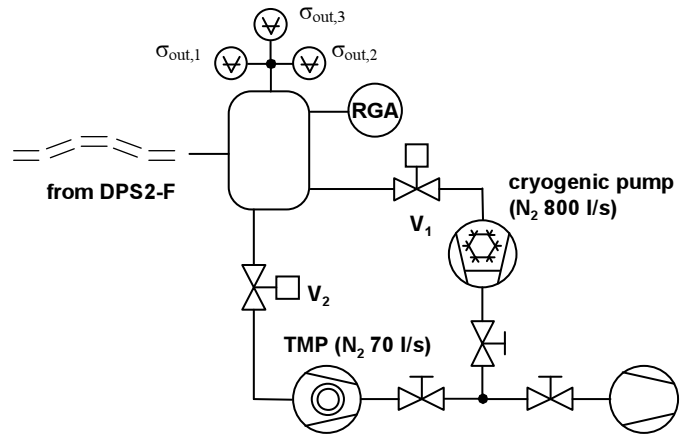}
\caption{Output gas-flow measurement system with the cryogenic pump and the RGA}
\label{measurement}
\end{figure}

The transmitted gas flow at the \mbox{DPS2-F} outlet was measured using the system schematically shown in Fig. \ref{measurement}. To provide vacuum conditions as close as possible to those at the boundary between the \mbox{DPS2-F} and the CPS, a cryogenic pump with pumping speed of 800~l/s for nitrogen is installed in the measurement system. A TMP with pumping speed of 70~l/s for nitrogen is installed for evacuation of the system, as well as for providing additional pumping capacity. Remote-controlled valves $\text{V}_1$ and $\text{V}_2$ close off the pumps from the central volume in different stages of operation. Since the output gas flow from the \mbox{DPS2-F} is in the order of $10^{-7} \; \text{mbar l/s}$, mass-selective measurement of the gas flow is mandatory, because outgassing originating from the last stage of the \mbox{DPS2-F} and from the measurement system itself is of the same order of magnitude or higher. For that purpose, the Pfeiffer QMS200 Prisma\textsuperscript{TM} residual gas analyzer (RGA) was calibrated for measurement of gas flow as described in Sec. \ref{sec-calibration}. 

For total pressure measurement and monitoring, two MKS Baratron\textsuperscript\textregistered transducers, one with a full scale of 1.33~mbar ($\upsigma_\text{out,1}$), and one with a full scale of 1333~mbar ($\upsigma_\text{out,2}$), as well as one cold-cathode gauge ($\upsigma_\text{out,3}$) were installed. 

\section{Molecular flow simulations and the analysis of the measurement}
\label{sec-analysis}

The goal of the present experiment can be expressed as follows: determine the ratio of tritium flow across the boundary surface $A_1$ between the WGTS and the \mbox{DPS2-F}, and the boundary surface $A_2$ between the \mbox{DPS2-F} and the CPS in the KATRIN operating conditions. In the measurement, however, the conditions of the KATRIN experiment are only approximated. One important difference is that test gases other than tritium are used in this measurement. Beside that, the measurement system and the KATRIN CPS have \textit{a priori} different pumping speeds. Another difference worth noting is that the beaming profile of the gas at the \mbox{DPS2-F} inlet is different in this measurement than in KATRIN conditions. Detailed information on the beaming profile of the gas coming from KATRIN WGTS is at the moment unfortunately not available for reasons of complexity of the WGTS outlet volume inlcuding six pumping ports \cite{Mal09}. However, we estimate that the beaming profile at the beginning of the \mbox{DPS2-F} beam tube does not significantly influence the beaming profile across the chamber of the first pumping port situated 1~m further downstream, and thus does not alter the throughput ratio between the first TMP and the second beam tube section. This was confirmed by a simulation comparing the throughput ratio in the case of the cosine angular distribution of the molecules at the inlet and in the case of a strongly forward-focused distribution, proportional to $\text{cos}^4 \theta$. The throughput ratio was the same to within the 1~\% statistical uncertainty of the simulation. Therefore it can be concluded that the beaming profile at the \mbox{DPS2-F} inlet does not significantly influence the final result for the reduction factor.

The following analysis concentrates on the effect of the different molecular masses of the gas flowing through the \mbox{DPS2-F} and on the difference in pumping speeds between the measurement system and the KATRIN CPS. These two considerations can be expressed as differences in the capture factors at the respective boundary surfaces of the \mbox{DPS2-F}. To determine the influence of the capture factors, $\alpha^{2800}_{TMP}$ of the MAG~W2800 TMP and $\alpha_{out}$ at the \mbox{DPS2-F} outlet, on the measured gas-flow reduction factor, the modified flow-matrix method described by X. Luo et al. in Ref. \cite{Luo06} was applied. In this method, matrix inversion technique is used in a similar way as in the well known view-factor method of solving stationary molecular flows, but with the difference that the transmission probabilities are considered not between all surface elements of the system, but only between the boundary surfaces, through which molecules can enter and/or leave the system, i.e. pumps and injection chambers. The transmission probabilities between the boundary surfaces were obtained by the Monte-Carlo simulation. 

The Molflow+ simulation code \cite{Ker09} was used to obtain the transmission matrix corresponding to the as-built geometry of the \mbox{DPS2-F} beam tube. Since a detailed description of the bellows in the beam tube would require a prohibitively large number of facets, all bellows were approximated by straight tube segments with radii corresponding to the minimum radii of the respective bellows. Hence the reductions in the middle of the beam-tube segments 2 to 5. The somewhat larger reduction in the middle of the first beam-tube segment corresponds to the FT-ICR housing that was installed at that place at the time of the experiment. The minimum radius of the bellows in the pumping ducts is equal to the inner radius of the duct itself. The entire geometry of the \mbox{DPS2-F} beam tube was described by a mesh consisting of 88292 facets. Due to the approximate mirror symmetry through the middle plane normal to the beam tube, it was sufficient to run three simulations to obtain the $6 \times 6$ matrix of transmission probabilities between the six boundary surfaces. In these three simulations, molecules were starting from the inlet, the first and the second pumping ports, respectively. The probabilities for molecules starting from the third and the fourth pumping port, as well as from the outlet, are then derived according to the mentioned symmetry. The number of simulated molecules was $1.70 \times 10^7$ for the simulation starting from the inlet, $1.39 \times 10^7$ starting from the first pumping port, and $7.6 \times 10^6$ molecules starting from the second pumping port.

Once the transmission matrix is known, the reduction factor can be easily obtained by solving it for any given set of capture factors $\alpha^{2800}_{TMP}$ and $\alpha_{out}$. The capture factors were obtained for all test gases by measurements and simulations described in the following sections.

\subsection{The capture factor of the TMP}

\begin{figure}
\centering
\includegraphics{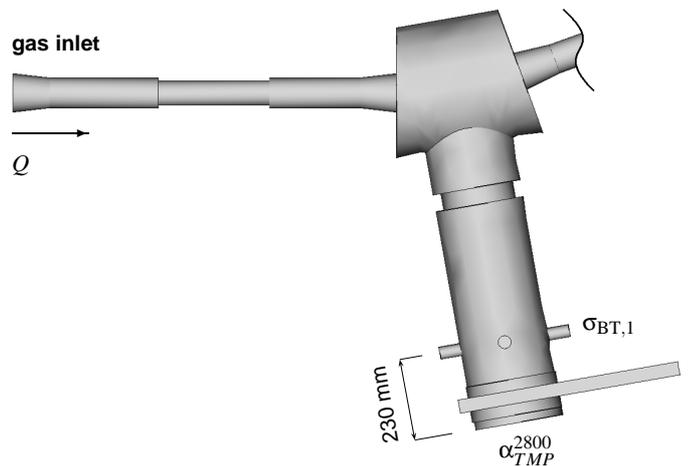}
\caption{Geometry of the inlet section of the beam tube with the first pumping port.}
\label{S-tester}
\end{figure}

To determine the capture factor of the installed TMP, the pressure $p$ in the pumping duct was measured for several different values of the injected gas flow $Q$. As the TMP capture factor depends on the mass of the gas species, this dependence was studied on DPS2-F for each of the five test gases. The correlation between the $Q/p$ ratio and the TMP capture factor were found using the simulation, according to the model shown in Fig. \ref{S-tester}. The model consists of 22783 facets. Three runs of the model were performed to obtain the transmission probabilities between the inlet $A_{in}$, the surface $A_\sigma$ where the pressure sensor was located, and the TMP flange $A_{TMP}$. The surface $A_\sigma$ is neither a pumping nor a desorption surface, but it was included in order to obtain the ratio between the number of hits $H_\sigma$ on the surface, and the total number of simulated moleculrs $D$, by solving the matrix equation. The $Q/p$ ratio is then calculated, assuming the Maxwell-Boltzmann distribution of the velocities of the molecules impinging onto $A_\sigma$, as

\begin{equation}
\label{qp}
Q/p = \sqrt{\frac{k_B T}{2 \pi m}} \frac{H_\sigma}{A_\sigma D}
\end{equation}

The number of the simulated molecules in each run was adjusted so that the relative statistical error of the smallest transmission probability calculated in the run is of the order of 1~\%. Thus in the run with the molecules starting from the inlet, $1.07 \times 10^7$ molecules were simulated, starting from the facet of the pressure sensor, $1.56 \times 10^6$ molecules, and $5.26 \times 10^6$ molecules starting from the pump.

In the measurement, only the pump in the first pumping port was pumping, and the gate valves of the remaining three TMP were closed, so that there was no net gas flow beyond the first pumping port. The pressure $p$ was measured in the upper CF40 flange in the pumping duct by a calibrated Baratron\textsuperscript{\textregistered} transducer denoted $\upsigma_\text{BT,1}$ in Fig. \ref{S-tester}. The resolution of the transducer was $10^{-6} \text{\ mbar}$ and the accuracy was 0.5~\%. The distance between the TMP flange and the center of the sensor flange was 23~cm. The gas was injected from the injection system via a manual leak valve and determined the throughput from the pressure drop over time in the injection buffer ($Q = V \frac{\Delta p_0}{\Delta t}$).\footnote{The necessary gas flow in this part of the measurement was higher than reachable by the installed gas-flow controller, and was therefore injected via a manual leak valve present in the system.} The pressure $p_0$ was measured by the transducer $\upsigma_\text{in,2}$. The absolute value of the relative pressure drop rate $\left | \frac{1}{p}\frac{\Delta p}{\Delta t} \right |$ never exceeded $1 \times 10^{-4} \ \text{s}^{-1}$. The accuracy of the determination of $Q$ was checked afterwards by direct comparison with a MKS type 179B mass-flow meter, and the results were in agreement to 1.5~\%, which is within the combined uncertainties of the pressure-drop method and the mass-flow meter. 

\begin{figure}
\centering
\includegraphics{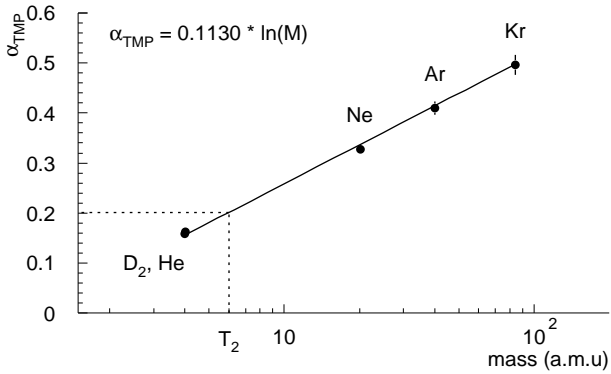}
\caption{Measured values of the capture factor of the \mbox{DPS2-F} TMP as a function of the molecular mass of the test gases. The line represents the fit of the function $\alpha^{2800}_{TMP} = a \text{\ ln}(M)$.}
\label{a-tmp}
\end{figure}

The results for $\alpha^{2800}_{TMP}$ are shown in Fig. \ref{a-tmp}. The uncertainties were derived from the uncertainties in the determination of $Q$ and $p$ due to the scattering of the pressure data. Note that $Q$ is particularly sensitive to the instabilities of the pressure signal $p_0$ in the injection buffer. Series of up to 300 automatically stored pressure values were used to minimize this uncertainty. The data taking rate was 1~s\textsuperscript{-1}.

The scaling of the capture factor of TMP with the logarithm of molecular mass observed in Ref. \cite{Mal07} is well reproduced in Fig. \ref{a-tmp}. The functional dependence $\alpha^{2800}_{TMP} = a \text{\ ln}(M)$ was fitted to the data by linear regression, and the fitted value $a = 0.1130 \pm 0.0024$ was obtained.\footnote{One should note that the necessary simplifications in the simulated geometry cause an in principle unknown systematic uncertainty in this value. We assume this uncertainty to be comparable or smaller than the uncertainty resulting from the uncertainty in pressure measurement} Using this value, the capture factor of 0.20 was derived for tritium. 

\subsection{The capture factor of the measurement system}

\begin{figure}
\centering
\includegraphics{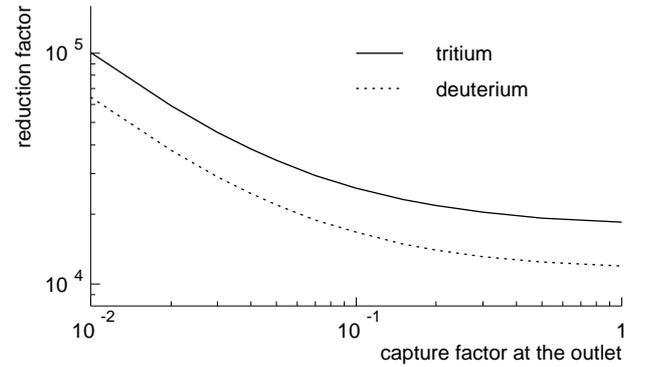}
\caption{Dependence of the gas-flow reduction factor on the capture factor at the outlet of the \mbox{DPS2-F}}
\label{cf-dependence}
\end{figure}

The dependence of the reduction factor on $\alpha_{out}$ for the final design of the \mbox{DPS2-F} as calculated by the matrix method is shown in Fig. \ref{cf-dependence} for the cases of tritium and deuterium. One can readily see that for capture factors below 0.1, the influence on the reduction factor can not be neglected. It is, thus, essential to estimate the capture factor of the CPS, as well as that of the measurement system, in order to make reliable predictions for KATRIN running conditions. 

In order to estimate the capture factor $\alpha_{ms}$ of the measurement system, the first step is to determine the capture factors $\alpha_{cryo}$ and $\alpha^{70}_{TMP}$ at the boundary surfaces between the central volume of the measurement system and the valves $\text{V}_1$ and $\text{V}_2$ leading to the cryogenic and the TMP, respectively (see Fig. \ref{measurement}). To determine $\alpha^{70}_{TMP}$ measurements of the ratio of the pressure $p$ to the gas flow $Q$ were performed for each of the test gases for the case when only the TMP is pumping. The correlation between the ratio $Q/p$ and the capture factor $\alpha^{70}_{TMP}$ was then found with the results of the flow simulation. In the case when both the TMP and the cryogenic pump are pumping, the pressure was too low to be measured by the capacitance-diaphragm gauge. Therefore the RGA was used to determine the ratio of the pressures for the given throughput when only the TMP is pumping, and when both the TMP and the cryogenic pump are pumping. The capture factor $\alpha_{cryo}$ was then found using the flow simulation. The estimated uncertainty of $\alpha_{cryo}$ is 20~\% due to the fluctuations in the RGA sensitivity (see Sec. \ref{sec-calibration}). 

Once $\alpha_{cryo}$ and $\alpha^{70}_{TMP}$ were determined, $\alpha_{ms}$ was obtained by gas-flow simulation. The gas flow was simulated using a model consisting of a 0.5~m long segment at the end of the beam tube of the \mbox{DPS2-F}, the thermal decoupler separating the beam tube and the end flange, as well as the measurement system. The last 0.5~m segment of the beam tube was included in order to properly reproduce the molecular beaming in that part of the volume. The capture factor $\alpha_{ms}$ is defined at the boundary surface at the end of the \mbox{DPS2-F} beam tube, in front of the thermal decoupler. The model consists of 10857 facets. The number of simulated molecules in these simulations was in the range between 30000 and 100000. The number of hits on the surface used to estimate the pressure was between 10000 and 60000, depending on whether only the TMP or both pumps were simulated, on the type of gas, as well as on the total number of simulated particles.

The results for $\alpha_{ms}$ for different test gases are listed in table \ref{capture-meas}. The uncertainties were derived from the uncertainty of $\alpha_{cryo}$ by variation of the parameter $\alpha_{cryo}$ in the simulation. The uncertainty of 20~\% in $\alpha_{cryo}$ results in an uncertainty of between 8~\% and 12~\% in $\alpha_{ms}$, depending on the gas.

\begin{table}
\caption{Capture factors $\alpha_{ms}$ of the measurement system for different test gases}
\label{capture-meas}
\begin{center}
\begin{tabular}{@{} l c @{}}
Gas type & $\alpha_{ms}$ \\
\hline
He & $0.077 \pm 0.009$ \\
$\text{D}_2$ & $0.099 \pm 0.012$\\
Ne & $0.111 \pm 0.009$\\
Ar & $0.113 \pm 0.010$ \\
Kr & $0.116 \pm 0.010$\\
\end{tabular}
\end{center}
\end{table}

\subsection{The capture factor of the KATRIN CPS inlet}

\begin{figure*}
\centering
\includegraphics[width=140mm]{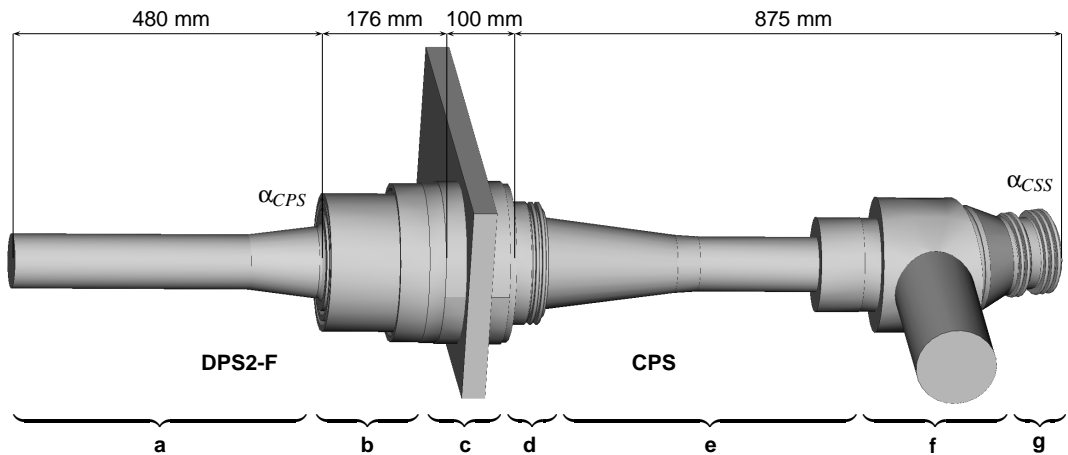}
\caption{3D model used to estimate the capture factor $\alpha_{CPS}$: a) the 0.5~m section at the end of the \mbox{DPS2-F} beam tube, b) the thermal decoupler of the \mbox{DPS2-F} beam tube, c) the gate valve, d) the thermal decoupler of the CPS inlet, e) the first section of the CPS beam tube, f) the regeneration pumping port, g) the bellow to the cryosorption segment. The approximate positions of the boundary surfaces where the capture factors $\alpha_{CPS}$ and $\alpha_{CSS}$ are defined are indicated by the corresponding symbols.}
\label{CPS}
\end{figure*}

Fig. \ref{CPS} shows the model of the vacuum volume of the CPS inlet, together with the last part of the \mbox{DPS2-F} that was used to estimate the capture factor $\alpha_{CPS}$. Again the last 0.5~m segment of the \mbox{DPS2-F} beam tube was included in order to properly reproduce the molecular beaming in that part of the volume. The end of this segment represents the end of the volume shown in Fig.\ref{DPS2}, and it is followed by the thermal decoupler of the \mbox{DPS2-F}, the gate valve, the thermal decoupler of the CPS and the CPS inlet, including the regeneration port. The simulated volume ends immediately before the cryosorption section of the CPS, represented by the boundary surface with the capture factor $\alpha_{CSS}$. The model consists of 47716 facets. The relatively large number of facets is due to the fact that the bellows were included in some detail. The number of simulated molecules was at least $1 \times 10^5$ in each of the simulated cases.

The capture factor $\alpha_{CPS}$ depends on the transmission probability $T$ of the volume consisting of the thermal decoupler of the \mbox{DPS2-F}, the gate valve, the thermal decoupler of the CPS and the CPS inlet until the cryosorption segment with capture factor $\alpha_{CSS}$. It is difficult to precisely predict $\alpha_{CSS}$ due to the lack of experimental data on the sticking coefficient for tritium on the argon cryosorption surface at 3~K. However, based on the simulation from Ref. \cite{Luo08}, with a conservative assumption for the sticking coefficient of 0.5, a reasonable lower limit for $\alpha_{CSS}$ can be set at 0.84.

Results for the extremal values of the capture factor $\alpha_{CSS}$ are listed in table \ref{capture-CPS}. It becomes clear that $\alpha_{CPS}$ is dominated by the transmission probability $T$ of the intermediate volume, and the dependence on $\alpha_{CSS}$ is rather weak. As both extremal values of $\alpha_{CSS}$ represent assumptions that are independent of the type of gas, and $T$ is gas-independent in the free-molecular regime, $\alpha_{CPS}$ is also independent of the type of gas.

\begin{table}
\caption{Capture factor of the CPS. Here $\alpha_{CSS}$ is the capture factor at the beginning of the cryosorption section and $\alpha_{CPS}$ is the capture factor at the end of the \mbox{DPS2-F} beam tube.}
\label{capture-CPS}
\begin{center}
\begin{tabular}{@{} l c c @{}}
 & $\alpha_{CSS}$ & $\alpha_{CPS}$ \\
\hline
Ideal cryosorption surface & 1.00 & 0.107 \\
Conservative estimate & 0.84 & 0.104 \\
\end{tabular}
\end{center}
\end{table}

\section{Calibration of the measurement system}
\label{sec-calibration}

As shown in Ref. \cite{Mal08}, residual gas analyzers can be used for reasonably accurate quantitative measurements of partial pressures in simple gas mixtures, provided that careful in-situ calibration is done. Beside that, in a system with a constant flow in the free-molecular regime, the partial pressure is in direct proportion to the throughput of the gas. This fact was used to calibrate the response of the RGA against the throughput of different test gases. The Baratron\textsuperscript\textregistered gauge $\upsigma_\text{out,1}$ (see Sec. \ref{sec-setup}) was used as the standard as described in the following sections. Although this calibration was not performed against a primary standard, the accuracy is sufficient for the present purpose. The obtained calibration parameters are valid for the flow measurement only with the geometry and the pumping capacity of the measurement system with which they were obtained. 

To perform the calibration, the outlet of the injection system was connected to the inlet of the flow-measurement system via a CF-250 gate valve of the same type as the valves at the inlet and the outlet of the \mbox{DPS2-F}. The gas was injected into the collection system via an orifice of 0.1~mm diameter drilled in a 0.7~mm thick stainless-steel disc installed in the place of one of the gaskets of the manual valve in one of output lines of the injection system (Fig. \ref{injection}). The orifice provided a fixed conductance for test gases in molecular flow from the injection into the measurement system. Further downstream the injection chamber shown in Fig. \ref{inlet} distributed the flow of the injected gas over the cross-sectional area similar to the flow conditions at the \mbox{DPS2-F} outlet.

\subsection{Conductance of the orifice}
\label{sec-orifice}

The conductance of the orifice was measured in the following way: the injection buffer was evacuated, and then filled with the test gas to a given pressure $p_0$ in the range between 0.08 and 0.6~mbar. The throughput through the orifice was determined by recording the pressure rise in the measurement system. The results for the gas load due to outgassing and leaks in the measurement system was corrected by measuring the pressure rise without gas injection. This correction was always below 10~\%. The relative fluctuations of repeated measurements with the same type of gas were below 5~\%. The measured conductance was independent of $p_0$ up to at least 1~mbar, for all gases. 

The conductance of the orifice for different gas types scales with the inverse square root of molecular mass,

\begin{equation}
\label{scaleConductance}
C_A = \sqrt{\frac{m_{He}}{m_A}} C_{He}
\end{equation}

\begin{figure}
\centering
\includegraphics{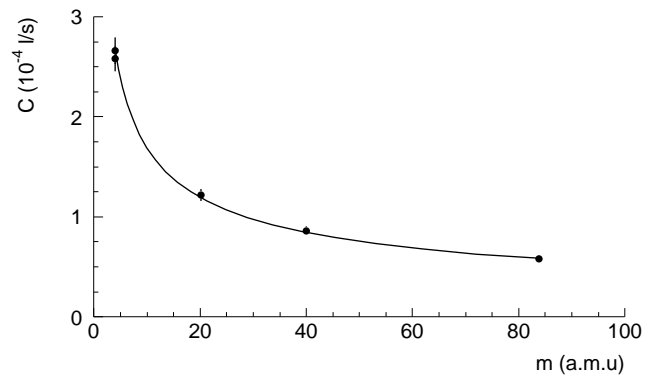}
\caption{Measured conductance of the calibration orifice as a function of the molecular mass of the gas. Eq. (\ref{scaleConductance}) was fitted to the data.}
\label{c-vs-m}
\end{figure}

Results for the measured conductance of the orifice are shown in Fig.~\ref{c-vs-m} as a function of the molecular mass of the respective test gases. Isotopic separation was neglected, and the average masses of natural isotopic mixtures were used. The line in Fig.~\ref{c-vs-m} represents the fit of the Eq. \eqref{scaleConductance} to the data. The only fitted parameter is the conductance for helium, and the resulting value is $C_{He} = (2.67 \pm 0.06) \times 10^{-4} \text{l/s}$.  

With this orifice, gas flows from $10^{-8}$ to $10^{-6} \text{mbar l/s}$ were generated by maintaining a constant pressure $p_0$ in the order of $10^{-4}$ to $10^{-2} \text{mbar}$ in the buffer vessel. To maintain $p_0$ in this range, the test gas was injected into the buffer via the mass-flow controller FC2 (Fig. \ref{injection}) while continuously pumping the buffer with the TMP. The pressure level in the buffer was controlled by adjusting the throughput through FC2. The flow $Q = p_0 C_A$ through the orifice was determined from the pressure reading $p_0$ of the Baratron\textsuperscript{\textregistered} gauge $\sigma_{in,1}$ and the known conductance $C_A$ of the orifice.

\subsection{RGA}

\begin{table}
\caption{Calibration factors of the mass spectrometer for the gas flow $Q$ into the measurement system and for the particle densities $n$. The relative uncertainty of the calibration factors is estimated to 20\% (see discussion in the text).}
\label{RGA-params}
\begin{center}
\begin{tabular}{@{} l c c c c @{}}
Gas type  & \multicolumn{2}{c}{$Q/I_{RGA} \left[ \frac{\text{mbar l/s}}{\text{A}} \right]$} 
          & \multicolumn{2}{c}{$n/I_{RGA} \left[ \frac{\text{m}^{-3}}{\text{A}} \right]$} \\
          & Filament 1 & Filament 2 & Filament 1 & Filament 2 \\
\hline
$\text{D}_2$ & $8.2 \times 10^5$ & $7.4 \times 10^5$ & $4.5 \times 10^{26}$ & $4.1 \times 10^{26}$ \\
He & $1.4 \times 10^6$  & $1.1 \times 10^6$  & $1.1 \times 10^{27}$ & $8.5 \times 10^{26}$  \\
Ne & $2.1 \times 10^6$  & $1.2 \times 10^6$  & $2.1 \times 10^{27}$ & $1.2 \times 10^{27}$  \\
Ar & $8.1 \times 10^5$  & $3.1 \times 10^5$  & $1.2 \times 10^{27}$ & $4.5 \times 10^{26}$  \\
Kr & $1.6 \times 10^6$  & $5.1 \times 10^5$  & $3.2 \times 10^{27}$ & $1.0 \times 10^{27}$  \\
\end{tabular}
\end{center}
\end{table}

To calibrate the RGA, the ion currents in the Faraday Cup were recorded for the available range of flow rates from $10^{-8}$ to $10^{-6} \text{mbar l/s}$ for each type of test gas, for each emission filament separately. Both the cryogenic and the TMP were active. All measurements showed excellent linearity. The resulting calibration factors are listed in Tab.~\ref{RGA-params}. In repeated measurements over three months, the calibration factors obtained for helium  were scattered about 20~\% around the mean value. This scatter might come from the fluctuations in the ionization and transmission probabilities of the RGA, and possibly also from fluctuations in the pumping speed of both pumps in the system. This scattering represents the dominant contribution to the uncertainty of the calibration parameters. The relative uncertainty of the orifice conductance is 2~\% (Sec. \ref{sec-orifice}), and the errors of the calibration parameter due to the scattering of the calibration data from a single measurement are typically between 1~\% and 2~\%. Thus the overall relative uncertainty of the calibration factors was estimated to be around 20~\%. 

For informational purpose, the ratio of the particle density $n$ at the position of the RGA ion source to the gas flow $Q$ was estimated by simulation, and the corresponding calibration parameters for the particle density were also determined. These parameters are valid for any chamber geometry with this type of RGA. However, it is important to note that these calibration parameters may be sensitive to the user-adjustable cathode and reference potentials of the ion source, as well as the potential of the extractor electrode. Therefore, great care was taken to ensure that these settings remain constant throughout the calibration and measurement work.

\section{Results}
\label{sec-results}

The measurement of the \mbox{DPS2-F} reduction factor was performed with the beam tube at room temperature. A range of settings on the mass-flow-controller between 0.1 and 1.1 sccm were used, corresponding to the flow rates 0.0018 to 0.020 mbar l/s for deuterium and 0.0025 to 0.028 mbar l/s for the noble gases at room temperature.\footnote{Note that the flow rates for the noble gases are obtained by applying appropriate correction factors, due to the different sensitivity of the thermal-transfer method of mass-flow measurement.}

The measured results $R_{ms}$ are listed in table \ref{R_meas}, together with estimated reduction factors $R_{CPS}$ for the test gases under KATRIN running conditions (CPS at the outlet). The latter values were obtained by correcting the measured values for the effect of the capture factor at the outlet (see Tabs. \ref{capture-meas} and \ref{capture-CPS} and Fig. \ref{cf-dependence}). The uncertainties for all values were derived from the relative uncertainties of the calibration factors and the uncertainties of the correction for the capture factor of the measurement system. The dominant contribution comes from the uncertainties of the calibration factors

\begin{table}
\caption{Reduction factors for the test gases: directly measured values and estimates for the standard KATRIN operating conditions (with the CPS at the outlet)}
\label{R_meas}
\begin{center}
\begin{tabular}{@{} l c c @{}}
Gas type & $R_{ms}$ & $R_{CPS}$ \\
\hline
He & $(1.86 \pm 0.37) \times 10^4$ & $(1.66 \pm 0.34) \times 10^4$ \\
D$_2$ & $(1.85 \pm 0.37) \times 10^4$ & $(1.81 \pm 0.37) \times 10^4$ \\
Ne & $(4.1 \pm 0.82) \times 10^4$ & $(4.1 \pm 0.83) \times 10^4$ \\
Ar & $(4.9 \pm 1.0) \times 10^4$ & $(4.9 \pm 1.0) \times 10^4$ \\
Kr & $(5.6 \pm 1.1) \times 10^4$ & $(5.7 \pm 1.1) \times 10^4$ \\
\end{tabular}
\end{center}
\end{table}

\begin{figure}
\centering
\includegraphics{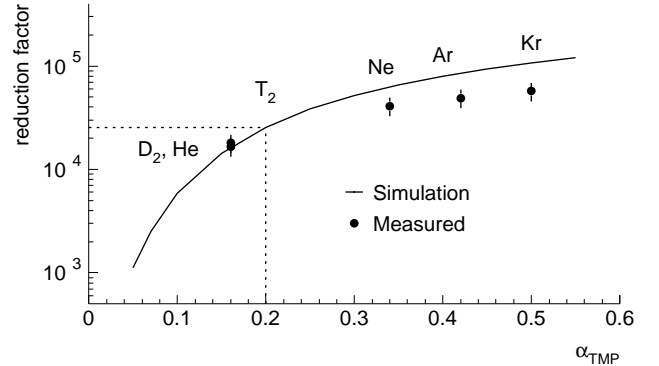}
\caption{The reduction factor as a function of the capture factor $\alpha^{2800}_{TMP}$ of the \mbox{DPS2-F} TMP for different gas species}
\label{fig-RF}
\end{figure}

The estimated reduction factors $R_{CPS}$ for all test gases are plotted against the corresponding TMP capture factors in Fig. \ref{fig-RF} in comparison with the simulation results. A systematic discrepancy of up to a factor 2 between the results of simulation and $R_{CPS}$ derived from the measurement is observed for heavier gas species. A possible source for this discrepancy might be in the necessary simplifications of the beam-tube geometry used in the simulations.

Using the value 0.2 for the capture factor of the TMP for tritium (see Fig. \ref{a-tmp}), one arrives at a value of $2.5 \times 10^4$ of the reduction factor for tritium. 

The \mbox{DPS2-F} beam tube was initially designed with a slightly different geometry than the present one. The main differences are  the beam-tube diameter and the geometry of the pumping ducts. In the production, the beam-tube diameter was redesigned from 75~mm to 86~mm in order to provide space for two \mbox{FT-ICR} ion traps \cite{Diaz09}, as well as for several electrostatic dipole electrodes \cite{Reim09, Wind11}, without affecting the transmission of the beta electrons. These instruments are important for the measurement of ion concentrations and for their removal from the KATRIN transport section. Moreover, due to mechanical design constraints, the pumping ducts had to be built ca. 300~mm longer than in the design geometry, and with an inner flange that reduces the diameter of the duct down to 204~mm at one point. Both these modifications result in a significant reduction of the duct conductance. 

Due to these changes, the present value of the reduction factor is four times lower than the design value of $1 \times 10^{5}$. The initial design value was estimated independently by calculation of conductances \cite{Bonn03}, and by molecular-flow simulations equivalent to those used in this work \cite{Luo06, KATRIN04}.  

The reduction factor is expected to improve after the installation of the \mbox{FT-ICR} ion traps and the dipole electrodes. The diameter of the remaining beam-tube sections can also be reduced to the original 75~mm by insertion of cylindrical placeholders. According to the gas flow simulations, the reduction factor will increase two times if the beam-tube diameter is reduced to 75~mm over the entire length. 

Measurement of the reduction factor will be repeated at the nominal operating temperature of the beam tube at 80~K, in order to investigate the influence of temperature on the reduction factor.

\section{Summary}
\label{sec-summary}

A system for a mass-selective measurement of gas-flow in the range of $10^{-7} \text{mbar l/s}$ at the output of the Differential Pumping Section \mbox{DPS2-F} of the KATRIN experiment was set up and calibrated. Using this system, the reduction factor for deuterium, helium, neon, argon and krypton was measured in the \mbox{DPS2-F} at room temperature. Molecular-flow simulations of the measurement system, the CPS inlet, and the pumping ports of the \mbox{DPS2-F} were performed. The measurement and simulation of the reduction factor were performed with only one \mbox{FT-ICR} housing installed in the first beam-tube section. The remainder of the beam tube was empty. The difference of the measured reduction factor with respect to the design value is due to the modifications in the beam tube geometry since the initial design, and can be partly recovered by reduction of the effective beam tube diameter. 

The measured value for gases with mass 4 is in excellent agreement with the simulated value. The estimated reduction factor for tritium is $2.5 \times 10^{4}$, based on the simulation using the value of the TMP capture factor for tritium that was obtained by interpolation between the measured values of the capture factor for other gases.

\section*{Acknowledgements}

The authors would like to thank the entire \mbox{DPS2-F} team from the Institute for Technical Physics (ITP), the Institute for Nuclear Physics (IK) and the tritium laboratory (TLK) at KIT. The group for Building Infrastructure and Experimental Technics of the Institute for Nuclear Physics (IK) at KIT, under the leadership of H. Krause and S. Horn, were responsible for the design of the necessary infrastructure and the assembly of the vacuum systems for the \mbox{DPS2-F}. Particular thanks are due to our colleagues J. Bonn, O. Kazachenko, N. Kernert, X. Luo and O. Malyshev for fruitful discussions on more than one occasion, as well as to J. Wolf for proof reading and for help with the technical presentation of the paper.

For the geometry definition in the flow simulations the FreeCAD open-source CAD software \cite{FCAD} was used.

This work has been funded by the Deutsche Forschungsgemeinschaft within the framework of the Transregional Collaborative Research Centre No. 27 ``Neutrinos and Beyond'', as well as by the German Ministry for Education and Research under project code 05CK5VKA/5. 

%\section*{References}

%\bibliographystyle{elsarticle-num}
\bibliography{RedFact}

\end{document}